\documentclass[english,prb,twocolumn,showpacs,superscriptaddress]{revtex4}
\usepackage[dvips]{epsfig}
\usepackage[T1]{fontenc}
\usepackage[latin9]{inputenc}
\usepackage{amsfonts}
\usepackage{bm}
\usepackage{bbm}
\usepackage{amsmath,amssymb,lscape,float}
\usepackage{graphicx}

\makeatletter
\@ifundefined{textcolor}{}
{%
 \definecolor{BLACK}{gray}{0}
 \definecolor{WHITE}{gray}{1}
 \definecolor{RED}{rgb}{1,0,0}
 \definecolor{GREEN}{rgb}{0,1,0}
 \definecolor{BLUE}{rgb}{0,0,1}
 \definecolor{CYAN}{cmyk}{1,0,0,0}
 \definecolor{MAGENTA}{cmyk}{0,1,0,0}
 \definecolor{YELLOW}{cmyk}{0,0,1,0}
 }

\makeatother

\usepackage{babel}
\begin{document}
\title{Transmon-based simulator of nonlocal electron-phonon coupling:\\
a platform for observing sharp small-polaron transitions}
\author{Vladimir M. Stojanovi\'c\footnote[2]{Electronic mail: stojanovic@physics.harvard.edu}}
\affiliation{Department of Physics, Harvard University, Cambridge, MA 02138, USA}
\author{Mihajlo Vanevi\'c}
\affiliation{Department of Physics, University of Belgrade, Studentski Trg 12, 11158 Belgrade, Serbia}
\author{Eugene Demler}
\affiliation{Department of Physics, Harvard University, Cambridge, MA 02138, USA}
\author{Lin Tian\footnote[3]{Electronic mail: ltian@ucmerced.edu}}
\affiliation{School of Natural Sciences, University of California, Merced, California 95343, USA}

\date{\today}

\begin{abstract}
We propose an analog superconducting quantum simulator for a one-dimensional model 
featuring momentum-dependent (nonlocal) electron-phonon couplings of 
Su-Schrieffer-Heeger and ``breathing-mode'' types. Because its corresponding 
coupling vertex function depends on both the electron- and phonon quasimomenta, 
this model does not belong to the realm of validity of the Gerlach-L\"{o}wen 
theorem that rules out any nonanalyticities in single-particle properties.
The superconducting circuit behind the proposed simulator entails an array of transmon qubits 
and microwave resonators. By applying microwave driving fields to the qubits, a
small-polaron Bloch state with an arbitrary quasimomentum can be prepared in this system 
within times several orders of magnitude shorter than the typical qubit decoherence times.
We demonstrate that -- by varying the externally-tunable parameters -- one can 
readily reach the critical coupling strength required for observing the sharp transition from 
a nondegenerate (single-particle) ground state corresponding to zero quasimomentum ($K_{\textrm{gs}}=0$) 
to a twofold-degenerate small-polaron ground state at nonzero quasimomenta $K_{\textrm{gs}}$ 
and $-K_{\textrm{gs}}$. Through exact numerical diagonalization of our effective Hamiltonian, 
we show how this nonanalyticity is reflected in the relevant single-particle properties 
(ground-state energy, quasiparticle residue, average number of phonons). We also show that the 
proposed setup provides an ideal testbed for studying the nonequilibrium dynamics of small-polaron 
formation in the presence of strongly momentum-dependent electron-phonon interactions.
\end{abstract}
\maketitle

\section{Introduction}\label{intro}
Based on the pioneering ideas of Feynman and Lloyd~\cite{feynmansim,Lloyd:96}, and bolstered 
by developments in technology and methods for manipulation and control, the field of quantum 
simulation is at the current frontier of physics research~\cite{Cirac+Zoller:12}. 
Its overarching goal is to help us understand the behavior of complex quantum many-body 
systems by studying their simpler, highly-controllable counterparts. The field has already 
matured enough to allow realizations of various quantum spin models, models with bosonic degrees
of freedom, and even of those that go beyond the conventional low-energy physics paradigm~\cite{Georgescu+:14}.
In particular, polaronic systems have quite recently attracted attention 
among researchers in the field, as evidenced by the proposals for simulating such systems with 
trapped ions~\cite{Stojanovic+:12,Mezzacapo+:12}, cold polar molecules~\cite{Herrera+Krems:11,Herrera+:13}, 
Rydberg atoms or ions~\cite{Hague+MacCormick}, and superconducting (SC) circuits~\cite{Mei+:13}.

Conceived by Landau and Pekar as a by-product of their investigation of charge carriers in polar 
semiconductors~\cite{landaupekar}, the polaron concept has played a pivotal role in studies of 
electron-phonon (e-ph) interaction ever since~\cite{AlexandrovDevreese}. It envisions an excess 
carrier (electron, hole) in a narrow-band semiconductor (or an insulator) strongly interacting 
with the host-crystal lattice vibrations. The effective mass of the carrier is increased 
-- compared to the bare-band value -- due to a self-induced lattice deformation  
that causes an effective ``dressing'' of the carrier by virtual phonons (self-trapping). 
Polaronic carriers have been found in materials ranging from amorphous 
semiconductors~\cite{Baily+Emin:06,Agarwal+:13} to colossal-magnetoresistive 
oxides~\cite{Adams+:00} to undoped cuprates~\cite{Roesch+:05}. More recently, polaronic 
behavior has been identified in cold-atomic systems~\cite{ColdPolaron}.
Besides, the generalized polaron concept includes almost any instance of a quantum 
particle strongly coupled to a bosonic environment, giving rise to 
a field-theoretic model of a fermion interacting with a scalar boson field.

In systems with short-range e-ph coupling, a typical size of the phonon 
cloud around a carrier does not exceed a unit cell of the host crystal (small polarons).
Such carriers are usually studied within the framework of the Holstein molecular-crystal 
model, a paradigm for the polaron-crossover problem~\cite{Holstein:59}. It
describes purely local -- hence momentum-independent -- interaction of 
tightly-bound electrons with dispersionless (Einstein) phonons. Owing to a large body of 
work over the past five decades~\cite{AlexandrovDevreese}, both static and dynamical 
properties of this model are well understood by now. Recent small-polaron studies have, 
however, focussed on strongly momentum-dependent e-ph interactions~\cite{Stojanovic+:04}. 
An important example is furnished by Su-Schrieffer-Heeger (SSH) coupling (also known 
as Peierls-type or off-diagonal coupling), which accounts for the dependence of 
electronic hopping integrals upon phonon states and has a significant bearing on transport 
properties of $\pi$-electron systems, such as organic semiconductors~\cite{PeierlsOrganics}, 
carbon nanotubes~\cite{SSHnanotubes}, and graphene-based nanostructures~\cite{StojanovicGraphene}. 
Such strongly momentum-dependent couplings, with vertex functions that depend both on the electron 
and phonon quasimomenta, are also relevant for fundamental reasons. Namely, they do not belong 
to the realm of validity of the Gerlach-L\"{o}wen theorem~\cite{Gerlach+Lowen:87}, 
which asserts that (single-particle) e-ph models generically show smooth dependence of the 
ground-state energy on the coupling strength. While this theorem
was long believed to be of quite general validity, it applies only to
momentum-independent (Holstein-like) couplings  and those that do depend on the 
phonon- but not on the electron quasimomentum. The latter are exemplified 
by the ``breathing-mode'' (BM) coupling~\cite{Slezak++:06}, relevant in the 
cuprates. 

The field of SC qubits~\cite{SCqubitReview} was revolutionized by the development of circuit 
quantum electrodynamics (circuit QED)~\cite{Wallraff++:04,Blais+:04,GirvinCQEDintro}, 
which allowed both fast quantum-gate realizations~\cite{DiCarlo++:09,StojanovicToffoli:12} 
and demonstrations of many quantum-optics effects. Quite recently, circuit QED has 
come to the fore as a versatile platform for on-chip analog quantum 
simulation~\cite{Houck+:12,Schmidt+Koch:13}. Spurred by some recent advances in 
the realm of SC quantum devices, especially the striking increase in achievable coherence 
times of transmon qubits (from $1\:\mu$s  to nearly $100\:\mu$s)~\cite{Paik++:11,Rigetti++:12},
theoretical proposals have already been set forth for simulating Bose-Hubbard-type 
models~\cite{Hohenadler+:12,Gangat+:13,VillalongaCorrea+:13}, coupled-spin-~\cite{Tian:10,Zhang++:14,LasHeras++:13}
and spin-boson models~\cite{Egger+Wilhelm:13,Kurcz+:13}, topological states
of matter~\cite{Schmidt+:13}, and gauge fields~\cite{Kapit:13}. 
Along similar lines, in this paper we propose an SC-circuit 
emulation of a one-dimensional model featuring momentum-dependent 
e-ph couplings of SSH and BM types. This analog simulator 
entails SC transmon qubits and microwave resonators, the standard building 
blocks of circuit-QED systems~\cite{Koch++:07}. The role of qubits in our 
system is to emulate spinless-fermion excitations, where the pseudospin-$1/2$ operators 
representing the qubits are mapped to fermionic ones through the Jordan-Wigner 
transformation. At the same time, the resonator modes (microwave photons) play 
the role of dispersionless (Einstein) phonons.  

We show that the suggested setup allows realization of strong e-ph coupling regime,
characterized by a small-polaron formation. Furthermore, it enables 
the observation of a sharp transition from a nondegenerate single-particle ground 
state at zero quasimomentum to a twofold-degenerate one corresponding to a pair
of equal and opposite (nonzero) quasimomenta.
To demonstrate the feasibility of our simulation scheme, we show that -- by 
applying appropriate pump pulses on the qubits -- the relevant small-polaron 
states can be prepared within time scales several orders of magnitude shorter 
than the relevant qubit decoherence times. 

The proposed simulator is particularly suitable for a detailed characterization 
of small-polaron ground states (or even excited states), by extracting their phonon 
content through direct counting of photons on the resonators. This unique tool, which is not 
at our disposal in traditional solid-state systems, also opens up the possibility to address
experimentally the nonequilibrium aspects of polaron physics, i.e., the complex problem of 
polaron-formation dynamics~\cite{Emin+Kriman:86,Ku+Trugman:07}. The question as to how rapidly upon 
creation (injection) of a single electron (hole), i.e., e-ph interaction quench, the cloud 
of correlated virtual phonons around it forms and results in a ``dressed'' polaronic quasiparticle is 
poorly understood at present. This dynamical process, which is expected to be 
particularly complex in systems with strongly momentum-dependent e-ph interactions,
has quite recently attracted considerable attention~\cite{Li+Movaghar+Ratner:13,Feng+:13}.

The remainder of this paper is organized as follows. The layout of the simulator 
circuit is presented in Sec.~\ref{circuit_setup}, together with the derivation 
of the effective Hamiltonian and discussion of the relevant parameter regime.
In Sec.~\ref{GSanalysis}, we first discuss the character of the momentum
dependence of the simulated e-ph coupling, then some technical aspects related to
exact numerical diagonalization, and finally, the results obtained for the small-polaron 
ground state. Special emphasis is placed on the occurrence of a sharp transition and the
ensuing nonanalyticities in relevant single-particle properties (ground-state energy,
quasiparticle residue, average number of phonons). Section \ref{StatePrepControl} 
starts with a brief description of our envisioned state-preparation protocol, 
followed by a discussion of the experimental-detection and robustness aspects 
of the simulator. In Sec.~\ref{Extract_via_Ramsey} we lay out the scheme for
extracting the relevant retarded Green's functions and the spectral function using
the many-body Ramsey interference protocol. In addition, we explain in detail how our 
setup can be used for studying the dynamics of polaron formation. We summarize and 
conclude in Sec.~\ref{Conclusions}.

\section{Analog simulator and its effective Hamiltonian}\label{circuit_setup}
\subsection{Circuit layout and underlying Hamiltonian} \label{circ_set_a}
\begin{figure}[t!]
\includegraphics[clip,width=8.6cm]{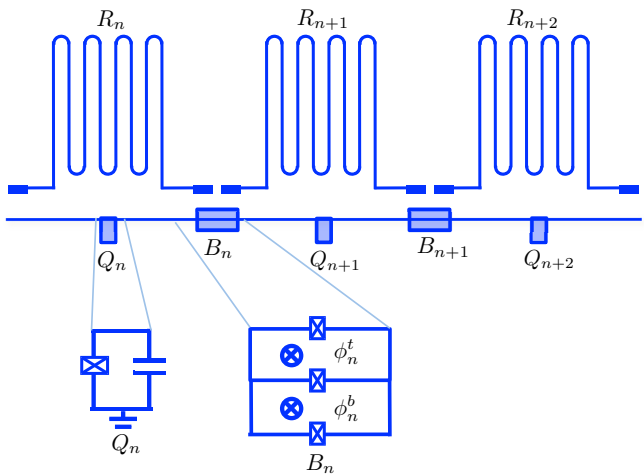}
\caption{\label{fig:circuit}(Color online) Schematic of the simulator circuit
containing transmon qubits $Q_{n}$, SC resonators $R_{n}$, 
and connecting circuits $B_{n}$ with three Josephson junctions. 
External magnetic fluxes $\phi_{n}^{b}$ and $\phi_{n}^{t}$ are threading 
the bottom- and top loops of each connecting circuit, respectively.
Note that the circuit elements are not drawn to scale.}
\end{figure}

A schematic of the SC circuit behind the simulator is shown in Fig.~\ref{fig:circuit}.
Each building block of the simulator consists of a transmon qubit (denoted as 
$Q_{n}$) and a SC resonator ($R_{n}$). The qubit emulates fermionic excitations, 
through mapping of its pseudospin-$1/2$ degree of freedom to spinless fermions via the 
Jordan-Wigner transformation. At the same time, the microwave photon modes of the SC resonator play
the role of Einstein phonons. In this system, adjacent qubits couple to each other via a connecting circuit 
denoted as $B_{n}$. Contrary to the more familiar circuit-QED setup, the resonators 
here do not couple directly to the qubits. Instead, the magnetic flux
of the resonator modes couples inductively to the connecting circuit
$B_{n}$, which enables a nearest-neighbor $XY$-type qubit-qubit coupling whose 
strength depends on the quantum dynamics of the resonator modes. 

The noninteracting Hamiltonian of the $n$-th repeating unit of the simulator,
containing qubit $Q_{n}$ and resonator $R_{n}$, can be written as 
\begin{equation}\label{eq:Hn}
H_{n}^{s}=\hbar\omega_{c}a_{n}^{\dagger}a_{n}
+\frac{E_{z}}{2}\:\sigma_{n}^{z}\:,
\end{equation}
where $E_{z}$ is the energy splitting of the qubit, and $\omega_{c}$
the frequency of the resonator mode; $a_{n}$ ($a^{\dagger}_{n}$) 
is the annihilation (creation) operator for the modes of the resonator 
$R_{n}$, while the Pauli matrix $\sigma_{n}^{z}$ represents the qubit $Q_{n}$. 

The connecting circuit $B_{n}$ consists of three Josephson junctions
and can be viewed as a generalized SQUID loop. Let $\varphi_{n}$
be the gauge-invariant phase variable of the SC island
of the $n$-th qubit, and $\varphi_{n}^{i}$ ($i=1,2,3$) the respective 
phase drops on the three junctions in the circuit $B_{n}$. Based
on flux-quantization rules~\cite{Devoret:97}, we then have
\begin{eqnarray}\label{eq:ph}
\varphi_{n}^{1} & = &\varphi_{n}-\varphi_{n+1}
+\frac{\phi_{n}^{t}}{2}\:,\nonumber\\
\varphi_{n}^{2} & = &\varphi_{n}-\varphi_{n+1}
-\frac{\phi_{n}^{t}}{2}\:,\label{eq:ph}\\
\varphi_{n}^{3} & = &\varphi_{n}-\varphi_{n+1}
-\frac{\phi_{n}^{t}}{2}-\phi_{n}^{b} \nonumber\:,
\end{eqnarray}
where $\phi_{n}^{t}$ and $\phi_{n}^{b}$ are the respective magnetic 
fluxes in the top- and bottom loops of $B_{n}$; both are expressed 
in units of $\Phi_{0}/2\pi$, where $\Phi_{0}\equiv hc/(2e)$ is the 
flux quantum.

The resonator modes $a_{n}$ and $a_{n+1}$ couple inductively to
the top loop of $B_{n}$ (see Fig.~\ref{fig:circuit}). 
The total magnetic flux in the top loop is given by 
\begin{equation}\label{eq:phin}
\phi_{n}^{t}=\pi\cos(\omega_{0}t)+\phi_{n,\textrm{res}}\:,
\end{equation}
where the first term is the flux from an external ac drive
(with amplitude $\pi$ and frequency $\omega_{0}$), while the 
second one
\begin{equation}\label{eq:dphin}
\phi_{n,\textrm{res}}=\theta_{n,n+1}(a_{n+1}+a_{n+1}^{\dagger})-
\theta_{n,n}(a_{n}+a_{n}^{\dagger})
\end{equation}
is the flux from the resonator modes. The coupling constants $\theta_{n,n'}$
quantify the contribution of the resonator modes $a_{n'}$ to the 
connecting circuit $B_{n}$. The sign of $\theta_{n,n+1}$ can be designed 
to be either the same or opposite to that of $\theta_{n,n}$, depending on 
the geometry of the qubit-resonator coupling. Here we choose 
$\theta_{n,n+1}=\theta_{n,n}\equiv\delta\theta$. In terms of the 
resonator parameters, $\delta\theta$ is given by~\cite{OrlandoDelinBook}  
\begin{equation}\label{eq:phinm}
\delta\theta=\frac{2e}{\hbar}\frac{A_{\textrm{eff}}}{d_{0}c}
\sqrt{\frac{\hbar\omega_{c}}{C_{0}}}\:,
\end{equation}
where $A_{\textrm{eff}}$ is an effective coupling area, $C_{0}$ the capacitance
of the resonator, and $d_{0}$ the effective spacing in the resonator.
In writing Eqs.~\eqref{eq:dphin} and \eqref{eq:phinm}, we assumed that the magnitudes
of fluxes from all resonators are equal and that the signs of the fluxes
from adjacent resonators are opposite to each other. 

The flux $\phi_{n}^{b}$ in the bottom loop includes an external ac driving 
and a tunable dc flux $\phi_{b0}$:
\begin{equation}\label{eq:phid}
\phi_{n}^{b}=-\frac{\pi}{2}\cos(\omega_{0}t)+\phi_{b0}\:.
\end{equation}
Note that the ac part of $\phi_{n}^{b}$ has the same frequency, but opposite 
sign of its amplitude compared to the ac part of $\phi_{n}^{t}$. We choose these 
amplitudes so that the phase drops $\varphi_n^3$ on the bottom junctions 
do not have an explicit time dependence. This is one of the crucial ingredients 
in the derivation of the effective Hamiltonian in Sec.~\ref{circ_set_b}.

The ac magnetic fluxes in this simulator can be realized by fabricating control wires 
that couple both to the top- and bottom loops, and applying a microwave pump to the wires. 
At the same time, the dc magnetic flux in the bottom loops can be implemented by applying
a dc current to a separate control wire designed near those loops. Such an
implementation should ensure that no significant dc bias couples to the top loops.
It should also be emphasized that the only tunable parameters in this circuit
are the dc bias $\phi_{b0}$ and the frequency of the ac drive $\omega_{0}$. 

By generalizing the standard expression for the effective Josephson energy of 
a SQUID loop (with two junctions, threaded by a magnetic flux)~\cite{MakhlinEtAl:01}, 
the total Josephson energy of the connecting circuit $B_{n}$ can be written as 
\begin{equation}\label{eq:HJ0}
H_{n}^{J}=-\sum_{i=1}^{3}E^{i}_{J}\cos\varphi_{n}^{i} \:,
\end{equation}
where $E^{i}_{J}$ are the Josephson energies of the three junctions. 
We will hereafter assume that $E^{1}_{J}=E^{2}_{J}\equiv E_{J}$ 
and $E^{3}_{J}=E_{Jb}$. 

\subsection{Effective Hamiltonian in the rotating frame of the drive} \label{circ_set_b}
Given the explicit time dependence originating from the driving terms,  
we resort to studying this system in the rotating frame of the drive,
i.e., adopt the interaction picture defined by a reference Hamiltonian 
\begin{equation}\label{Hodef}
H_{0}=\hbar\omega_{0}\sum_{n}a_{n}^{\dagger}a_{n}\:.
\end{equation}
The system dynamics in this rotating frame are described by 
the Hamiltonian
\begin{equation}\label{HI_def}
H_{I}=e^{\frac{i}{\hbar}H_0t}\left[\sum_n (H_{n}^{s}+H_{n}^{J})
-H_0\right]e^{-\frac{i}{\hbar}H_0t}\:,
\end{equation}
which can be recast as
\begin{equation}\label{HI_simp}
H_{I}=\sum_n\left(\hbar\delta\omega a_{n}^{\dagger}a_{n}
+\frac{E_{z}}{2}\sigma_{n}^{z}\right)+e^{\frac{i}{\hbar}H_0t}
\sum_n H_{n}^{J}\:e^{-\frac{i}{\hbar}H_0t}\:,
\end{equation}
with $\delta\omega\equiv\omega_c-\omega_0$. The explicit form of the 
Josephson-coupling term $H_{n}^{J}$ in the rotating frame is derived in 
detail in Appendix~\ref{rotframe_deriv}. Its time-independent part is given by
\begin{equation}\label{eq:HJn_final}
\bar{H}_{n}^{J}=-2\left[t_{r}-\frac{1}{2}E_{J}J_{1}(\pi/2)
\phi_{n,\textrm{res}}\right]\cos(\varphi_{n}-\varphi_{n+1})\:,
\end{equation}
where $J_n(x)$ are Bessel functions of the first kind, and 
\begin{equation}\label{eq:tr0}
t_{r}=E_{J}J_{0}\left(\pi/2\right)
\left(1+\cos\phi_{b0}\right) 
\end{equation}
is determined by the Josephson energy of the bottom 
junction, which, for convenience, is chosen as $E_{Jb}=2E_{J}J_{0}(\pi/2)$.
Along with the first two terms in Eq.~\eqref{HI_simp}, 
the time-independent part $\sum_n\bar{H}_{n}^{J}$ of the Josephson-coupling 
term forms the effective Hamiltonian of the system in 
the rotating frame. Namely, the remaining (time-dependent) 
part of this transformed Josephson-coupling term can be neglected due to its 
rapidly-oscillating character (for details, see Appendix A), in accordance 
with the rotating-wave approximation (RWA).

By expanding $\cos(\varphi_{n}-\varphi_{n+1})$ to quadratic order in the phase 
difference $\varphi_{n}-\varphi_{n+1}$ and defining pseudospin-$1/2$ operators 
$\sigma_{n}$ that correspond to the lowest two eigenstates of the transmon, 
we find that
\begin{multline}\label{eq:cos}
-2\cos(\varphi_{n}-\varphi_{n+1})\approx -2+4\delta\varphi_{0}^{2}\\
-2\delta\varphi_{0}^{2}
\left(\sigma_{n}^{+}\sigma_{n+1}^{-}+\sigma_{n}^{-}\sigma_{n+1}^{+}
-\frac{\sigma_{n}^{z}+\sigma_{n+1}^{z}}{2}\right)\:. 
\end{multline}
Here $\delta\varphi_{0}^{2}\equiv (2E_{C1}/E_{J1})^{1/2}$ is the quantum displacement 
of the gauge-invariant phase ($E_{C1}$ and $E_{J1}$ are the charging- and
Josephson energies of an individual transmon, respectively); for
a typical transmon ($E_{J1}/E_{C1}\sim 100$), we have $\delta\varphi_{0}^{2}\approx 0.15$. 
Note that the above expansion in powers of $\varphi_{n}-\varphi_{n+1}$
is justified not only by the smallness of this phase difference in the regime of interest
for transmons ($E_{J1}\gg E_{C1}$), but also by the rapidly vanishing coefficients in 
the expansion (proportional to higher powers of $\delta\varphi_{0}^{2}$).
The full expression for $\bar{H}_{n}^{J}$ (not shown here) can easily be obtained 
by combining Eqs.~\eqref{eq:HJn_final} and \eqref{eq:cos}.

It should be stressed that in writing Eq.~\eqref{eq:cos} we have 
omitted the terms $\sigma_n^- \sigma_{n+1}^- + \sigma_n^+ \sigma_{n+1}^+$ .
Even in the most general (multilevel) treatment of transmons, such terms
are conventionally neglected by virtue of the RWA. 


We now switch to the spinless-fermion representation of the pseudospin-$1/2$ 
operators via the Jordan-Wigner transformation
\begin{equation}
1+\sigma_{n}^{z}\rightarrow 2c_{n}^{\dagger}c_{n} \quad,\quad 
\sigma_{n}^{+}\sigma_{n+1}^{-}+\sigma_{n}^{-}\sigma_{n+1}^{+}
\rightarrow c_{n}^{\dagger}c_{n+1}+\text{h.c.}\:.
\end{equation}
The effective Hamiltonian of the system in the rotating frame 
\begin{equation}\label{eq:Htr}
H_{\textrm{eff}}=H_{\textrm{ph}}+H_{\textrm{e}}+H_{\textrm{e-ph}}
\end{equation}
includes the free-phonon term with the effective phonon frequency $\delta\omega$,
\begin{equation}\label{freephonon}
H_{\textrm{ph}}=\hbar\delta\omega\sum_{n}a_{n}^{\dagger}a_{n}\:,
\end{equation}
the (spinless-fermion) excitation hopping term
\begin{equation}
H_{\textrm{e}} = -t_{0}\sum_{n}(c_{n}^{\dagger}c_{n+1}+\textrm{h.c.}) \:,
\end{equation}
with $t_{0}\equiv 2\delta\varphi_{0}^{2}\:t_r$ being the effective bare hopping 
energy, and the excitation-phonon coupling term $H_{\textrm{e-ph}}$ whose
explicit form will be specified shortly.
Note that, strictly speaking, $H_{\textrm{e}}$ also contains the diagonal (on-site 
energy) terms $c_{n}^{\dagger}c_{n}$ for spinless fermions, originating from the 
$\sigma_{n}^{z}$ terms in Eqs.~\eqref{eq:Hn} and \eqref{eq:cos}. Yet, in consistency 
with the usual practice in studying coupled e-ph models, we hereafter disregard them 
as they only represent a constant band offset for our itinerant fermionic excitations.

The coupling Hamiltonian $H_{\mathrm{e-ph}}$ consists of two contributions: 
the SSH term
\begin{multline}
H_{\mathrm{SSH}}=g\hbar\delta\omega\sum_{n}(c_{n}^{\dagger}
c_{n+1}+\textrm{h.c.}) \\
\times\left[\left(a_{n+1}+a_{n+1}^{\dagger}\right)
-\left(a_{n}+a_{n}^{\dagger}\right)\right] \:,
\end{multline}
and the BM term
\begin{multline}
H_{\mathrm{BM}}=-g\hbar\delta\omega\sum_{n}c_{n}^{\dagger}c_{n} \\
\times\left[\left(a_{n+1}+a_{n+1}^{\dagger}\right)-\left(a_{n-1}
+a_{n-1}^{\dagger}\right)\right]\:,
\end{multline}
where the dimensionless coupling strength $g$ is defined by the relation
\begin{equation}\label{gomega}
g\hbar\delta\omega=\delta\varphi_{0}^{2}\:E_{J}
J_1(\pi/2)\delta\theta\:.
\end{equation}
The SSH term physically accounts for the dynamical dependence of 
the hopping integral (i.e., the excitation bandwidth) on the phonon 
displacements $u_n \propto a_{n}+a_{n}^{\dagger}$, to first order in 
these displacements; it is nonlocal in that the hopping integral
between sites $n$ and $n+1$ depends on the displacements
on both sites~\cite{Stojanovic+:04}. The BM term, on the other hand, 
also describes a nonlocal e-ph interaction; it accounts for the antisymmetric 
coupling of the excitation density at site $n$ with the phonon displacements 
on the neighboring sites $n-1$ and $n+1$. Being  
an example of a ``density-displacement'' type coupling, it can be 
viewed as a nonlocal generalization of the Holstein-type e-ph interaction.

It is worthwhile to mention that an example of a real electronic system where 
both e-ph coupling mechanisms discussed here play important roles is furnished 
by the cuprates. In those materials, both mechanisms involve the planar Cu-O bond-stretching 
phonon modes, also known as breathing modes. These bond-stretching modes 
couple to electrons both via a direct modulation of the hopping integral
(SSH-type coupling)~\cite{Johnston+:10}, and through electrostatic changes 
in the Madelung energies that originate from displacements of orbitals 
(BM-type coupling)~\cite{Radovic+:08,ButkoEtAl:09}. 

\subsection{Relevant parameter range}
A suitable set of parameters for the SC resonators is given by $d_{0}=25\:\mu m$,
$A_{\textrm{eff}}=100\:\mu m^{2}$, $C_{0}=2$ fF, $\omega_{c}/2\pi=24$ GHz, while
the effective phonon frequency can be $\delta\omega/2\pi=200$ MHz or $300$ MHz. 
In addition, we choose $E_{J}$ for the Josephson junctions in the connecting circuits such that 
$2\delta\varphi_{0}^{2}E_{J}/2\pi\hbar=200$ GHz. From this choice of parameter values,
it follows that $\delta\theta=3.5\times10^{-3}$ and $g\delta\omega/2\pi=198$ MHz. 

\begin{figure}[t!]
\includegraphics[clip,width=0.47\textwidth]{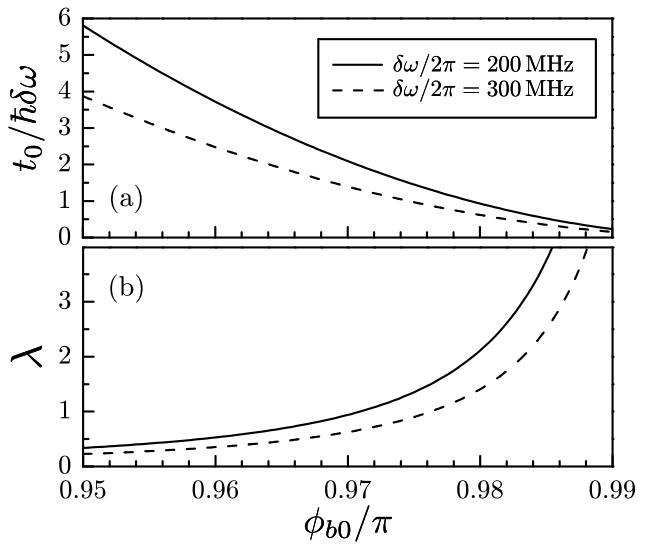}
\caption{\label{fig:para}Dimensionless parameters of the simulator:
(a) Ratio of the hopping integral and the phonon energy, and (b)
effective coupling strength $\lambda$, both shown as a function 
of the tunable dc-flux $\phi_{b0}$, for two different values of 
the effective phonon frequency $\delta\omega$.}
\end{figure}

The hopping integral $t_{0}$ can be adjusted in-situ by varying the dc 
magnetic flux $\phi_{b0}$ between the bottom two junctions. This is 
illustrated in Fig.~\ref{fig:para}(a), where the ratio $t_{0}/\hbar\delta\omega$ 
is shown for $\phi_{b0}/\pi$ between $0.95$ and $0.99$. 
For the chosen values of $\phi_{b0}/\pi$ we are mainly in the adiabatic 
regime ($t_{0}>\hbar\delta\omega$), entering the antiadiabatic regime 
($t_{0}<\hbar\delta\omega$) for $\phi_{b0}/\pi\approx 0.98$. For the 
same range of values for $\phi_{b0}$, the effective coupling strength
\begin{equation}\label{def_lambda}
\lambda=2g^{2}\frac{\hbar\delta\omega}{t_{0}}
\end{equation}
varies from the weak-coupling to the strong-coupling regime, 
as can be inferred from Fig.~\ref{fig:para}(b). 
For instance, with $\delta\omega/2\pi=200$ MHz and $\phi_{b0}/\pi=0.95$, 
we have $\lambda=0.34$; for the same value of $\delta\omega$, 
$\phi_{b0}/2\pi=0.98$ yields $\lambda=2.1$.

\section{Small-polaron signatures and sharp transition}\label{GSanalysis}
The eigenstates of our effective single-particle Hamiltonian $H_{\textrm{eff}}$
are the joint eigenstates of $H_{\textrm{eff}}$ and the total quasimomentum operator
$K_{\mathrm{tot}}=\sum_{k}k\:c_{k}^{\dagger}c_{k}+\sum_{q}q\:a_{q}^{\dagger}a_{q}$, 
where $c_{k}$ and $a_{q}$ are the excitation- and phonon annihilation 
operators in momentum space. For convenience, we will express quasimomenta in 
units of the inverse lattice period, so that $c_{k}=N^{-1/2}\sum_{n}e^{ikn}c_{n}$ and 
$a_{q}=N^{-1/2}\sum_{n}e^{iqn}a_{n}$. The eigenvalues of the operator 
$K_{\mathrm{tot}}$ will hereafter be denoted as $K$. In particular, $K_{\textrm{gs}}$
will stand for the quasimomentum corresponding to the ground state of the system 
(i.e., the dispersion minimum of the effective, dressed-excitation Bloch band). 

\subsection{Momentum dependence of the resulting e-ph coupling}\label{mom_dependence}
Before embarking on a numerical calculation of the ground state of the system,
it is worthwhile to analyze the momentum dependence of the vertex function 
corresponding to the resulting e-ph coupling term:
\begin{equation}\label{mscoupling}
H_{\mathrm{e-ph}}=N^{-1/2}\sum_{k,q}\gamma(k,q)\:
c_{k+q}^{\dagger}c_{k}(a_{-q}^{\dagger}+a_{q}) \:.
\end{equation}
This vertex function 
\begin{equation}\label{vertex_total}
\gamma(k,q)=2ig\hbar\delta\omega\:\Big[\sin k+\sin q-\sin(k+q)\Big] 
\end{equation}
consists of the $k$- and $q$-dependent SSH part
$\gamma_{\mathrm{SSH}}(k,q)=2ig\hbar\delta\omega\:[\sin k-\sin(k+q)]$, 
and the BM part $\gamma_{\mathrm{BM}}(q)=2ig\hbar\delta\omega\sin q$
that only features $q$ dependence. Since the overall vertex function
$\gamma(k,q)$ depends on both $k$ and $q$, it does not belong to the
domain of applicability of the Gerlach-L\"{o}wen theorem~\cite{Gerlach+Lowen:87}, 
which rules out a nonanalytic behavior of the single-particle
quantities. While $k$- and $q$-dependent couplings do not necessarilly
lead to nonanalyticities~\cite{Alvermann:07}, such nonanalyticity indeed 
occurs in a model with the pure SSH coupling [vertex function 
$\gamma_{\mathrm{SSH}}(k,q)$]~\cite{SSHnonanalytic}.
In the following (see Sec.~\ref{sharp_trans}), we show that the model with 
combined SSH and BM couplings studied here displays
a similar sharp transition between a quasifree excitation and a small polaron.

\subsection{Details of exact diagonalization}
To determine the ground-state properties of our resulting coupled e-ph model, 
we employ the conventional Lanczos diagonalization~\cite{CullumWilloughbyBook} 
in combination with a controlled truncation of the phonon Hilbert space. 

The Hilbert space of the system is spanned by states given as direct
products $|n\rangle_e \otimes |\mathbf{m}\rangle_\text{ph}$.
Here, $|n\rangle_e=c_{n}^{\dagger}|0\rangle_e$ is the state of the excitation 
localized at the site $n$, $\mathbf{m}=(m_1,\ldots,m_N)$ are the phonon occupation
numbers, and $|\mathbf{m}\rangle_\text{ph} = \prod_{i=1}^N
(1/\sqrt{m_i!})(b_i^\dagger)^{m_i}|0\rangle_\text{ph}$. Restricting ourselves to 
a truncated phonon Hilbert space that includes states with at most $M$ phonons 
(total number on a lattice with $N$ sites), we take into account all $m$-phonon states
with $0\le m_i \le m$, where $m=\sum_{i=1}^N m_i \le M$. The dimension of the total 
Hilbert space is $D = D_\text{e} \times D_\text{ph}$, where $D_\text{e} = N$ and
$D_\text{ph}=(M+N)!/(M!N!)$.

To further reduce the dimension of the Hamiltonian matrix to be diagonalized, 
we exploit the discrete translational invariance of our finite system, 
mathematically expressed as the commutation $[H_{\textrm{eff}},K_{\mathrm{tot}}]=0$
of the Hamiltonian $H_{\textrm{eff}}$ and the total quasimomentum operator $K_{\mathrm{tot}}$.
This allows us to perform diagonalization of $H_{\textrm{eff}}$ in sectors corresponding
to the eigensubspaces of $K_{\mathrm{tot}}$. For that to accomplish, we make 
use of the symmetrized basis  
\begin{equation}
|K,\mathbf{m}\rangle = N^{-1/2} \sum_{n=1}^N e^{iKn}\,
T_n(|1\rangle_\text{e} \otimes |\mathbf{m}\rangle_\text{ph})\:,
\end{equation}
where $T_{n}$ denotes (discrete) translation operators. Thus, the dimension
of each $K$-sector of the total Hilbert space is $D_{K}=D_{\textrm{ph}}$. 

Following an established phonon Hilbert-space truncation procedure~\cite{Wellein+Fehske:97}, 
the system size ($N$) and the maximum number of phonons retained ($M$) are increased until
the convergence for the ground-state energy $E^{(M)}_{\textrm{gs}}$ and the phonon distribution 
is reached. Our adopted covergence criterion is that the relative error in the ground-state 
energy and the phonon distribution upon further increase of $N$ and $M$ is not larger than 
$10^{-4}$. While for the momentum-independent (completely local in real space) 
Holstein coupling the system size is essentially inconsequential, 
for nonlocal couplings of the kind studied here this is not the case.
In particular, the adopted quantitative convergence criterion is satisfied 
for the system size $N=10$ (with periodic boundary conditions) and requires 
the total of $M=8$ phonons. 

\subsection{Results and Discussion}\label{sharp_trans}
We now discuss the results obtained by exact diagonalization of our effective 
model. Unlike more conventional situation, in which the effective coupling
strength $\lambda$ [cf. Eq.~\eqref{def_lambda}] is changed by varying the 
dimensionless coupling constant $g$ (for fixed ratio of the relevant 
hopping integral and the phonon energy)~\cite{Herrera+:13}, 
here we work with a fixed value of $g\delta\omega$ [recall Eq.~\eqref{gomega}].
We effectively change $\lambda$ by varying the bare hopping integral $t_0$ 
through the experimentally-tunable parameter $\phi_{b0}$, in accordance 
with Eq.~\eqref{eq:tr0}.

\begin{figure}[t!]
\includegraphics[clip,width=0.47\textwidth]{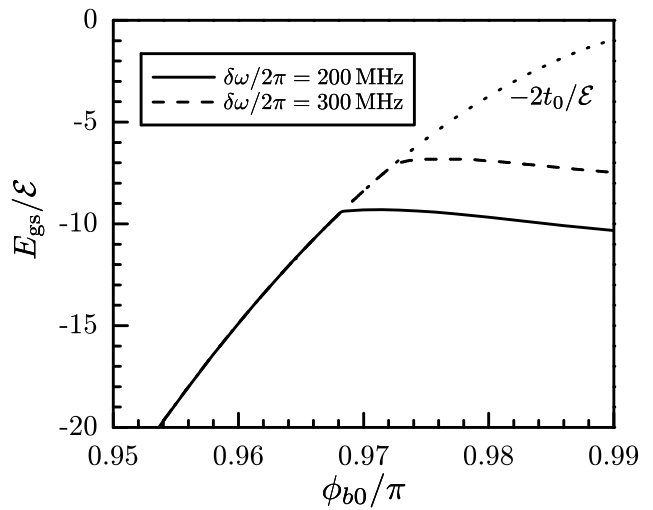}
\caption{\label{fig:E_GS}Ground-state energy, expressed in 
units of $\mathcal{E}\equiv\:10^{-3}\:\delta\varphi_{0}^{2}
\:E_J=2\pi\hbar\times 100$ MHz, as a function of the 
experimentally-tunable parameter $\phi_{b0}$.
The solid curve coresponds to the effective phonon frequency 
$\delta\omega/2\pi=200$ MHz, while the dashed curve corresponds 
to $\delta\omega/2\pi=300$ MHz.}
\end{figure}

Our main finding is that at a critical value of $\phi_b$
(i.e., of the effective coupling strength $\lambda$),
there is a sharp transition (nonanalyticity) of all relevant
quantities~\cite{SSHnonanalytic}. This transition originates 
from a (real) level crossing and is of first order.
It is illustrated in Fig.~\ref{fig:E_GS}, where the ground-state energy 
(expressed in units of $\mathcal{E}\equiv\:10^{-3}\:\delta\varphi_{0}^{2}
\:E_J=2\pi\hbar\times 100$ MHz) is shown as a function of $\phi_{b0}$. 
The sharp transition physically corresponds to a change -- at a critical value 
of $\phi_{b0}$ -- from a non-degenerate ground state that corresponds to the 
zero quasimomentum ($K_{\textrm{gs}}=0$), to a twofold-degenerate one 
corresponding to equal and opposite (nonzero) quasimomenta $K_{\textrm{gs}}$ 
and $-K_{\textrm{gs}}$. For $\delta\omega/2\pi=200$ MHz this critical value is 
$(\phi_{b0})_c\approx 0.968\:\pi$, while for $\delta\omega/2\pi=300$ MHz
we find $(\phi_{b0})_c\approx 0.972\:\pi$. The corresponding critical values 
of the effective coupling strength are $\lambda_c\approx 0.83$ 
and $0.72$, respectively.
\begin{figure}[b!]
\includegraphics[clip,width=0.47\textwidth]{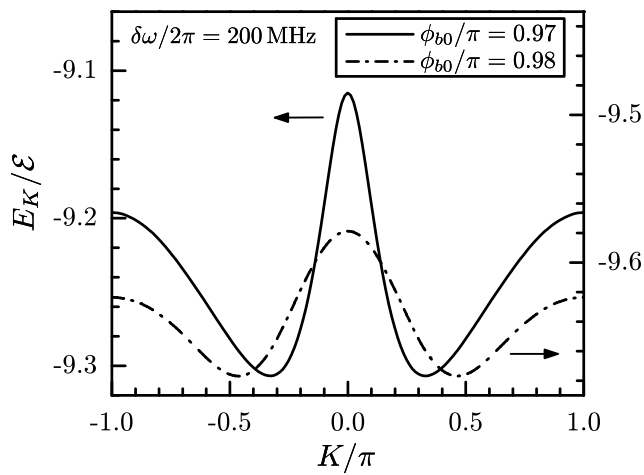}
\caption{\label{fig:E_K}Small-polaron Bloch dispersions for two different 
values of the tunable parameter $\phi_{b0}$, shown throughout the 
Brillouin zone. The solid curve coresponds to the $y$-axis scale 
marked on the left, while the dash-dotted curve corresponds 
to the scale on the right, as indicated by the arrows.}
\end{figure}
Shown in Fig.~\ref{fig:E_K} are two small-polaron Bloch-band dispersions throughout 
the Brillouin zone (for $\delta\omega/2\pi=200$ MHz and two different values 
of $\phi_{b0}$), both with band minima (ground states) at nonzero quasimomenta. 

For sufficiently strong coupling -- e.g., $\phi_{b0}\gtrsim 0.98$ for 
$\delta\omega/2\pi=200$ MHz -- the quasimomentum 
$K_{\textrm{gs}}$ corresponding to the single-particle ground state saturates 
at around $\pi/2$ (see the dash-dotted curve in Fig.~\ref{fig:E_K}).
It should be stressed that -- while the ground-state undergoes a sharp 
transition -- the quasimomentum $K_{\textrm{gs}}$ itself
varies smoothly between $K_{\textrm{gs}}=0$ and this saturation value
as $\phi_{b0}$ is increased beyond its critical value.  

Apart from the occurrence of a sharp transition, another interesting 
aspect of our findings is an effective ``compensation'' of SSH 
and BM couplings below the critical value of $\phi_{b0}$. This is indicated
in Fig.~\ref{fig:E_GS}, where the ground-state energy curve essentially follows 
the bare-dispersion curve $E=-2t_0(\phi_{b0})$ all the way up to the critical 
value of $\phi_{b0}$. This peculiar effect can be ascribed to 
the character of the resulting momentum dependence of the e-ph vertex 
function in Eq.~\eqref{vertex_total}, being a consequence of the fact that 
here the SSH and BM coupling strengths are the same~\cite{Stojanovic+:04}. 
This phenomenon could have profound consequences for
transport properties of real electronic systems with competing SSH and
BM couplings.
\begin{figure}[t!]
\includegraphics[clip,width=0.47\textwidth]{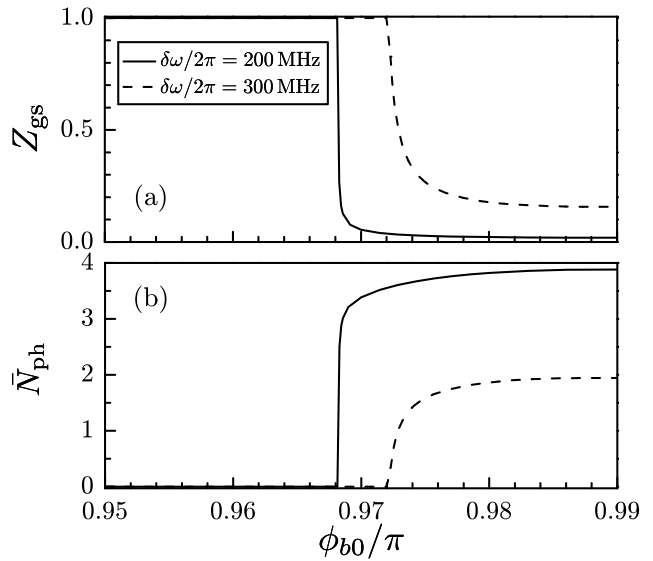}
\caption{\label{fig:ZgsNph} Characterization of the sharp transiton between 
a quasifree excitation and a small polaron: (a) Ground-state quasiparticle 
residue $Z_{\textrm{gs}}\equiv Z_{k=K_{\textrm{gs}}}$, and (b) average 
number of phonons $\bar{N}_{\text{ph}}$, versus the experimentally-tunable 
parameter $\phi_{b0}$. The solid and dashed curves corespond to 
$\delta\omega/2\pi=200$ MHz and $\delta\omega/2\pi=300$ MHz, 
respectively.}
\end{figure}

The central quantity for characterizing the small-polaron regime is the
quasiparticle residue $Z_{k}\equiv|\langle\Psi_{k}|\psi_{k}\rangle|^{2}$, 
i.e., the module squared of the overlap between the bare-excitation Bloch state 
$|\Psi_{k}\rangle\equiv c^{\dagger}_{k}|0\rangle$ and the (dressed) Bloch state 
$|\psi_{k}\rangle$ of the coupled e-ph system that corresponds to the same 
quasimomentum ($K=k$). In particular, having determined the ground-state wave 
function $\vert{\psi}_{\textrm{gs}}\rangle\equiv\vert{\psi}_{K=K_{\textrm{gs}}}\rangle$
we can compute $Z_{\textrm{gs}}\equiv Z_{k=K_{\textrm{gs}}}$, 
a quantity characterizing the ground state of the system. While $Z_{\textrm{gs}}\approx 1$  
indicates the weak-coupling regime (quasifree excitation), its 
reduced values in the strong-coupling regime [see Fig.~\ref{fig:ZgsNph}(a)] 
signify the presence of small polarons, with these two regimes 
being separated by a nonanalyticity at the critical value of $\phi_{b0}$. 
It is interesting to note that, unlike for the Holstein model where 
$Z_{\textrm{gs}}\approx 0$ for strong enough coupling,
here $Z_{\textrm{gs}}$ may saturate at a finite value. As can be inferred 
from Fig.~\ref{fig:ZgsNph}(a), for $\delta\omega/2\pi=300$ MHz we find 
such saturation at $Z_{\textrm{gs}}\approx 0.15$.

Another relevant quantity is the average number of phonons in 
the ground state
\begin{equation} \label{phave}
\bar{N}_{\text{ph}}\equiv{\langle{\psi}_{\textrm{gs}}\vert}
\:\sum_{n=1}^{N}a^{\dagger}_{n}a_{n}\:{\vert{\psi}_{\textrm{gs}}\rangle}\:. 
\end{equation}
The change of this quantity from values close to zero [see Fig.~\ref{fig:ZgsNph}(b)]
to a nonzero value $\bar{N}_{\text{ph}}\gtrsim 3$ marks the transition from
a quasifree excitation to a small polaron. The aforementioned 
effective compensation of the SSH and BM couplings below $(\phi_{b0})_c$
is reflected in the vanishing phonon-dressing of fermionic excitations,
the flat parts of the curves in Fig.~\ref{fig:ZgsNph}(b).

\begin{figure}[t!]
\includegraphics[clip,width=0.47\textwidth]{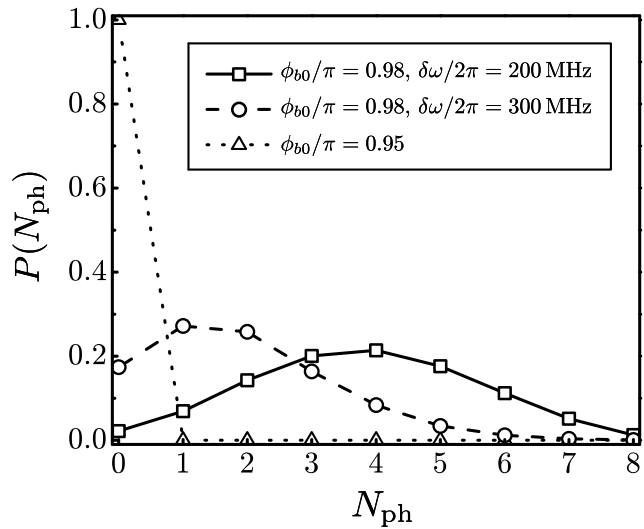}
\caption{\label{fig:PhDistr} Illustration of the multiphononic nature of small-polaron
excitations: Full ground-state phonon distribution at $\phi_{b0}/\pi=0.98$, 
for $\delta\omega/2\pi=200$ MHz (squares) and $\delta\omega/2\pi=300$ MHz (circles).
The dotted curve coresponds to the phonon distribution for $\phi_{b0}/\pi=0.95$, 
i.e., below the critical coupling strength.}
\end{figure}

The average phonon number $\bar{N}_{\text{ph}}$ is amenable to a direct measurement 
in our system, through measurements of photon numbers on different resonators (see Sec.~\ref{Detection}). 
Likewise, the second moment of the effective phonon distribution can also be extracted by
measuring the photon squeezing in the resonators~\cite{Marthaler+:08,Mei+:13}. More generally,
the complex multiphononic nature of small-polaron excitations can be fully captured 
by computing the entire phonon distribution, which is depicted in Fig.~\ref{fig:PhDistr} 
for both values of the effective phonon frequency $\delta\omega$ used above. 
While for $\delta\omega/2\pi=200$ MHz the distribution has a broad maximum at 
$N_{\textrm{ph}}\approx 4$, the one for $\delta\omega/2\pi=300$ MHz -- corresponding 
to a smaller effective coupling strength $\lambda$ [cf. Eqs.~\eqref{gomega} and 
\eqref{def_lambda}] -- has a weakly-pronounced maximum at around $N_{\textrm{ph}}\approx 1$.
By contrast, the dotted curve in Fig.~\ref{fig:PhDistr}, representing a typical 
phonon distribution for couplings below the critical one, is very strongly peaked
at $N_{\textrm{ph}}=0$. 

While phonon distributions peaked at zero phonons are characteristic of 
all small-polaron models in the weak-coupling regime (e.g., below the onset
of the polaron crossover in the Holstein model), the peculiarity of our findings 
is that here such distribution persists all the way up to the critical coupling
strength. Although the sharp transitions of the type found here do not result from 
any kind of cooperative behavior (as is typical for quantum phase transitions in many-particle 
systems), the observed peculiar behavior allows us to treat $\bar{N}_{\text{ph}}$ 
as an effective ``order parameter'' for the predicted sharp transition. Given that 
$\bar{N}_{\text{ph}}$ is a directly measurable quantity in our system, this fact 
will facilitate experimental verification of the existence of this transition. 

\section{State preparation, detection, and robustness}\label{StatePrepControl}
\subsection{State-preparation protocol}\label{StatePrep}
The feasibility of our simulation scheme is contingent upon the
ability to prepare the desired small-polaron Bloch states. For this purpose,
we make use of the state-preparation protocol proposed in our previous work 
(Ref.~\onlinecite{Mei+:13}), which is only briefly explained in the following. 

We assume that the initial state of the system is the vacuum state
$|G_{0}\rangle\equiv|0\rangle_{\textrm{e}}\otimes|0\rangle_{\textrm{ph}}$.
This state, with no excitations (i.e., with all qubits in their spin-down states) 
and all resonators in their respective vacuum states, can be prepared via thermalization 
in a low-temperature environment. Our target state is the dressed-excitation (in the 
special case, small polaron) Bloch state $\left|\psi_{K}\right\rangle$, corresponding 
to the eigenvalue $K$ of the total quasimomentum operator $K_{\mathrm{tot}}$.

The microwave driving required for preparing this state is envisioned 
to be of the form
\begin{equation}\label{driveop}
\Omega_q(t)=\frac{\hbar\beta(t)}{\sqrt{N}}\sum_{n}\left
(\sigma_{n}^{+}e^{-iqn}+\sigma_{n}^{-}e^{iqn}\right) \:,
\end{equation}
where $\beta(t)$ describes its time dependence, and the phase factors $e^{\pm iqn}$ 
indicate that spin-flip operations are applied to different qubits with a $q$-dependent 
phase difference. It is easy to show that  
the transition matrix element of the operator $\Omega_q(t)$ between 
the states $\left|G_{0}\right\rangle$ and $\left|\psi_{K}\right\rangle$ is 
given by~\cite{Mei+:13}
\begin{equation}\label{eq:Oqkt}
\left|\left\langle\psi_{K}\left|\Omega_q(t)\right|G_{0}
\right\rangle\right|=\hbar|\beta(t)|\sqrt{Z_K}\:\delta_{q,K} \:.
\end{equation}
Assuming that $\beta(t)=2\beta_{p}\cos(\omega_{K}t)$, 
where $\hbar\omega_{K}$ is the energy difference between the states
$\left|G_{0}\right\rangle$ and $\left|\psi_{K}\right\rangle$, in the RWA 
the two states are Rabi-coupled with the effective Rabi frequency 
$\beta_{p}\sqrt{Z_K}$. Thus, starting from the vacuum state $\left|G_{0}\right\rangle$, 
the state $\left|\psi_{K}\right\rangle$ will be prepared within a time interval 
of duration~\cite{Mei+:13} 
\begin{equation}
\tau_{\textrm{prep}}=\frac{\pi\hbar}
{2\beta_{p}\sqrt{Z_K}} \:.
\end{equation}
The form of the last expression is consistent with the expectation that the 
more strongly-dressed states (smaller $Z_K$) require longer preparation times. 
For the small-polaron ground states with $K=\pm K_{\textrm{gs}}$
these preparation times are much shorter than the decoherence time $T_2$. For 
instance, assuming that the pumping amplitude is
$\beta_{p}/(2\pi\hbar)=40$\:MHz and taking the values for $Z_{\textrm{gs}}$ at
the onset of the small-polaron regime [see Fig.~\ref{fig:ZgsNph}(a)], we respectively
find that $\tau_{\textrm{prep}}\approx 24$\:ns for $\delta\omega/2\pi=200$\:MHz, and 
$\tau_{\textrm{prep}}\approx 17$\:ns for $\delta\omega/2\pi=300$\:MHz.
The obtained state-preparation times are three orders of magnitude shorter than 
currently achievable decoherence times $T_2\sim 20-100\:\mu$s of transmon 
qubits~\cite{Paik++:11,Rigetti++:12}, this being a strong indication of the 
feasibility of our proposed protocol.

The above Rabi-coupling state-preparation protocol, which ensures energy and 
momentum conservation~\cite{Mei+:13}, can in principle be adapted to other  
systems by taking into account their underlying symmetries. Importantly, 
the ultimate sucess of this scheme in different systems will depend on the 
character of their underlying absorption spectra, as described by the 
corresponding spectral functions (see Sec.~\ref{poldyn_subsec}). As far as our system is 
concerned, it should be stressed that the absorption spectra of electronic systems coupled with 
optical phonons (i.e., phonon modes with a gap in their spectrum) are characterized by 
generic spectral functions in which the ``coherent'' part (with a finite spectral weight) 
is energetically well separated from the incoherent background~\cite{Engelsberg+Schrieffer:63,AlexandrovDevreese}.
This form of absorption spectra should here allow one to avoid inadvertent
population of other (excited) states while preparing a desired small-polaron 
ground state of the system.
 
\subsection{Experimental detection}\label{Detection}
Here we discuss the method for measuring the average phonon number 
$\bar{N}_{\textrm{ph}}$ [cf. Eq.~\eqref{phave}], a quantity which can be thought 
of as an order parameter for the sharp small-polaron transition (cf. Sec.~\ref{sharp_trans}). 
As already stressed above, in our implementation $\bar{N}_{\textrm{ph}}$ corresponds 
to the average total photon number on the resonators. Owing to the discrete 
translational symmetry of the system, the measurement of $\bar{N}_{\textrm{ph}}$ 
can be reduced to the measurement of the mean photon number of one 
of the resonators. This can be accomplished by adding an ancilla qubit  
which couples to this particular resonator~\cite{Mei+:13}, but only during 
the measurement; this qubit is assumed to be far-detuned from the resonator modes. 
By measuring the qubit state, the mean photon number on this resonator 
can be extracted, which multiplied with the total system size $N$ yields
the result for $\bar{N}_{\textrm{ph}}$. 

\subsection{Robustness of the simulator}\label{Robustness}
As in every other quantum-computation device, decoherence effects should
also be present in our simulator. Possible excitations in this system 
correspond either to flipping of the qubit states or to displacement of the
resonator modes, which are subject to the decoherence
of the qubits and resonators. In addition to very long dephasing times ($T_2\sim 20-100\:\mu$s)
of transmon qubits achieved in recent years, for coplanar waveguide resonators 
the damping time of the microwave photons can reach the same order of magnitude 
as $T_2$, with a quality factor larger than $10^6$. The relevant energy scales
in our simulator (effective phonon frequency, e-ph coupling strength, 
and hopping energy) are all of the order of $100$\:MHz, far exceeding the 
decoherence rates. Besides, as shown in Sec.~\ref{StatePrep}, even for very 
strongly-dressed polaron states the typical duration of the state-preparation 
protocol is three orders of magnitude smaller than the decoherence times. Finally, 
in the low-temperature environment of our system thermal excitations -- which 
here have energies of a few GHz -- can be safely neglected. 

The pump pulses of the kind described in Sec.~\ref{StatePrep} may, in principle, induce 
unwanted transitions (leakage) to higher energy levels in a transmon qubit~\cite{Gambetta+:11}. 
Namely, in all qubits based on weakly-anharmonic oscillators leakage from 
the two-dimensional qubit Hilbert space (computational states) is the leading 
source of errors at short pulse times~\cite{Fazio+:99}. This is especially pronounced 
if the pulse bandwidth is comparable to the anharmonicity.
However, in a typical transmon with $E_{J1}/E_{C1}\sim 50-100$ 
and a negative anharmonicity of $\sim 3-5\:\%$,
even pulses with durations of only a few ns are sufficiently frequency 
selective that the unwanted transitions can be neglected~\cite{GirvinCQEDintro}.
In our system, there is an off resonance of around $500$ MHz for such 
transitions. For a typical driving amplitude $\beta_p/2\pi \approx 40$\:MHz, 
the probability of leakage is below one percent, which is a reasonable 
error rate for the simulator.

Fabrication-related variations in the parameters of Josephson junctions, 
which are $1-2\:\%$ at best and typically up to $5\:\%$, are unavoidable  
even in modern-day SC circuits. This is, however, not expected to jeoperdize 
the observation of the predicted nonanalytic behavior in our system. Namely, as 
is known from general polaron theory (e.g., the Gerlach-L\"{o}wen theorem) 
the presence of such nonanalyticities (or, more often, the absence thereof) in a 
particular coupled e-ph model is intimately related to the type of 
the momentum dependence of the resulting coupling. This momentum dependence, in its
own right, depends on the character of the particular coupling mechanisms 
involved (recall discussions in the introduction and Sec.~\ref{mom_dependence}), 
rather than on quantitative details such as the magnitude of the bare hopping integral 
(which in our case is determined by $E_J$). Thus the occurrence of the real level 
crossing which gives rise to the sharp feature in the ground-state energy of the system 
should be quite robust; the small variations in parameters such as $E_J$ can 
only lead to a slight shift of the critical coupling strength at which this crossing 
takes place. That this shift is expected to be small -- for small variations $\Delta E_J$
of the Josephson energy ($\Delta E_J/E_J \lesssim 5\:\%$) -- can be inferred from the 
expression for the effective coupling strength $\lambda$ [cf. Eq.~\eqref{def_lambda}].
Because $\lambda \propto g^{2}/t_{0}$, where both $g$ and $t_0$ depend  
linearly on $E_J$, we have that $\Delta\lambda/\lambda=\Delta E_J/E_J$.

\section{Extracting correlation functions using a Ramsey 
sequence: study of the polaron-formation dynamics}\label{Extract_via_Ramsey}
\subsection{Relevant Green's functions} \label{GF_subsec}
Generally speaking, dynamical response functions -- Fourier transforms of retarded 
two-time Green's functions -- provide a natural framework for characterizing excitations 
in many-body systems. The relevant single-particle retarded two-time Green's 
function in the problem at hand is given by
\begin{equation}\label{anticommGF}
G_{+}^{\textrm{R}}(k,t)=-\frac{i}{\hbar}\:\theta(t)\langle\textrm{G}_0|
[c_k^{\dagger}(t),c_k]_{+}|\textrm{G}_0\rangle \:,
\end{equation}
where $c_k^{\dagger}(t)$ is a single-particle operator in the Heisenberg 
representation and $[\ldots]_{+}$ stands for an anticommutator.
More explicitly, $c_k(t)=U^{\dagger}_H(t)c_k U_H(t)$, where $U_H(t)$
is the time-evolution operator corresponding to the lab-frame counterpart $H=H(t)$
of our effective Hamiltonian $H_{\textrm{eff}}$ in the rotating frame. The explicit 
forms of $H$ and $U_H$ are not relevant for our present purposes. In fact,
the only relevant property of the Hamiltonian $H$ is that its symmetries in 
the spinless-fermion (pseudospin) sector of the problem are the same as those of 
the Hamiltonian $H_{\textrm{eff}}$, since this part of $H_{\textrm{eff}}$ preserves 
its form in the interaction picture.

It should be emphasized that, while the natural Green's functions for spinless
fermions are those that involve anticommutators [cf. Eq.~\eqref{anticommGF}], in the 
single-particle problem at hand the commutator Green's function
\begin{equation}\label{commGF}
G_{-}^{\textrm{R}}(k,t)=-\frac{i}{\hbar}\:\theta(t)\langle\textrm{G}_0|
[c_k^{\dagger}(t),c_k]_{-}|\textrm{G}_0\rangle \:.
\end{equation}
contains the same physical information. Namely, given that $|\textrm{G}_0\rangle$
is a vacuum state, we have that $c_k^{\dagger}(t)c_k|\textrm{G}_0\rangle=0$, which 
implies that in this special case $G_{-}^{\textrm{R}}(k,t)=-G_{+}^{\textrm{R}}(k,t)$.

Anticipating the use of a real-space experimental probe (see Sec.~\ref{Ramsey_subsec}), 
we further note that the last momentum-space Green's function can be retrieved from the 
real-space resolved ones, i.e., 
\begin{equation}
G^{\textrm{R}}_{-}(k,t)=N^{-1}\sum_{n,n'} e^{ik\cdot (n-n')}
G_{nn'}^{\textrm{R}}(t) \:,
\end{equation}
where $G_{nn'}^{\textrm{R}}(t)\equiv -(i/\hbar)\theta(t)\langle
\textrm{G}_0|[c_n^{\dagger}(t),c_{n'}]_{-}|\textrm{G}_0\rangle$.
By switching to the pseudospin-$1/2$ operators and taking into account
that the Jordan-Wigner string operators act trivially on the ground 
state $|\textrm{G}_0\rangle$, these real-space commutator Green's 
function can be rewritten as 
\begin{equation}
G_{nn'}^{\textrm{R}}(t)= -\frac{i}{\hbar}\:\theta(t)\langle\textrm{G}_0|
[\sigma_n^{+}(t),\sigma_{n'}^{-}]_{-}|\textrm{G}_0\rangle \:.
\end{equation}
They can further be transformed to the form
\begin{equation}\label{gnRt}
G_{nn'}^{\textrm{R}}(t)= \mathcal{G}^{xx}_{nn'}+\mathcal{G}^{yy}_{nn'}
-i(\mathcal{G}^{xy}_{nn'}-\mathcal{G}^{yx}_{nn'})\:,
\end{equation}
where 
\begin{equation}
\mathcal{G}^{\alpha\beta}_{nn'}\equiv -\frac{i}{\hbar}\:\theta(t)\langle\textrm{G}_0|
[\sigma_n^{\alpha}(t),\sigma_{n'}^{\beta}]_{-}|\textrm{G}_0\rangle 
\qquad (\:\alpha,\beta=x,y\:)
\end{equation}
and, for simplicity, we suppressed the time argument and 
the superscript $R$ in the notation for these Green's functions. 

\subsection{Many-body Ramsey interference protocol}\label{Ramsey_subsec}
The Ramsey-interference protocol is in principle applicable in any system where single-site 
addressability is available and yields naturally the real-space and time-resolved commutator Green's 
functions of spin operators~\cite{DeChiara+:08,Knap++:13}. In the problem at hand it involves the 
pseudospin degree of freedom of the transmon qubits. 

The general Rabi pulses can be parameterized as
\begin{equation}
R_n(\theta,\phi)\equiv\mathbbm{1}_{2\times 2}\cos\frac{\theta}{2} + 
i(\sigma_n^{x}\cos\phi-\sigma_n^{y}\sin\phi)\sin\frac{\theta}{2} \:,
\end{equation}
where $\theta=\Omega\tau$, with $\Omega$ being the Rabi frequency and $\tau$
the pulse duration; $\phi$ is the phase of the laser field. The Ramsey protocol
makes use of the special case $R_n(\phi)\equiv R_n(\theta=\pi/2,\phi)$ 
of such pulses, with $\theta=\pi/2$ and arbitrary $\phi$.

Quite generally, the Ramsey-interference protocol entails the following steps: 
(1) perform local $\pi/2$-rotation at site $n$ (with $\phi=\phi_1$);
(2) evolve the system during time $t$;
(3) perform local $\pi/2$-rotation at site $n'$, or global $\pi/2$-rotation (with $\phi=\phi_2$);
(4) measure the system in the $\sigma_z$ basis at site $n'$. In our system,
the described protocol leads to the measurement result given by the expectation value
\begin{multline}
M_{nn'}(\phi_1,\phi_2,t)=\\
\langle\textrm{G}_0|R_n^{\dagger}(\phi_1)
 U^{\dagger}_H(t) R_{n'}^{\dagger}(\phi_2) \sigma_{n'}^{z} R_{n'}(\phi_2)
 U_H(t) R_n(\phi_1)|\textrm{G}_0\rangle\:.
\end{multline}

The procedure for extracting relevant Green's function can be simplified 
by exploiting the symmetries of our system in the pseudospin sector.
Since its Hamiltonian involves a sum of an $XY$-coupling term and $\sigma_{n}^{z}$ 
terms [recall Eq.~\eqref{eq:cos}], our system has a $U(1)$ symmetry under $z$-axis 
pseudospin rotations, implying that $\mathcal{G}^{xx}_{nn'}=\mathcal{G}^{yy}_{nn'}$ 
and $\mathcal{G}^{xy}_{nn'}+\mathcal{G}^{yx}_{nn'}=0$. Another symmetry of our system 
is that under reflections with respect to the $z$ axis ($\sigma_n^{x}\rightarrow -\sigma_n^{x}$,
$\sigma_n^{y}\rightarrow -\sigma_n^{y}$, $\sigma_n^{z}\rightarrow \sigma_n^{z}$),
which implies that any expectation value involving an odd (total) number of
$\sigma_{n}^{x}$ and $\sigma_{n}^{y}$ operators is equal to zero. For a system
with these two symmetries, the Ramsey protocol measures~\cite{Knap++:13} 
\begin{multline}
M_{nn'}(\phi_1,\phi_2,t)=-\frac{1}{4}\big[\sin(\phi_1-\phi_2)
(\mathcal{G}^{xx}_{nn'}+\mathcal{G}^{yy}_{nn'}) 
\\
-\cos(\phi_1-\phi_2)(\mathcal{G}^{xy}_{nn'}
-\mathcal{G}^{yx}_{nn'})\big] \:.
\end{multline}
Thus the combinations $\mathcal{G}^{xx}_{nn'}+\mathcal{G}^{yy}_{nn'}$ and 
$\mathcal{G}^{xy}_{nn'}-\mathcal{G}^{yx}_{nn'}$ needed to recover 
$G_{nn'}^{\textrm{R}}(t)$ [recall Eq.~\eqref{gnRt}] can be obtained by
choosing the angles $\phi_1,\phi_2$ such that $\phi_1-\phi_2=\pm \pi/2$ and
$\phi_1=\phi_2$, respectively.

In the realm of SC qubits, the Ramsey-interference protocol is conventionally
used to determine the decoherence time $T_2$ of a single qubit, a procedure
known as the Ramsey-fringe experiment~\cite{Rigetti++:12}. The use of this protocol
is also envisioned for other types of manipulation, such as interaction-free 
measurements~\cite{Paraoanu+:06}. In the present work, we propose its use on 
pairs of qubits in a multi-qubit system, for the purpose of extracting the desired
two-time correlation- and Green's functions.

\subsection{Spectral function and its relation to the dynamics of small-polaron formation}\label{poldyn_subsec}
Provided that the single-particle retarded two-time Green's function [cf. Eq.~\eqref{anticommGF}]
is extracted as explained in Secs.~\ref{GF_subsec} and \ref{Ramsey_subsec}, we can also
obtain the information about the spectral properties of the system.
Namely, by Fourier-transforming this Green's function to 
momentum-frequency space, the spectral function is obtained as
$A(k,\omega)=-(1/\pi)\textrm{Im}\:G_{+}^{\textrm{R}}(k,\omega)$.
The spectral function can quite generally be represented in the form
\begin{equation}
A(k,\omega)=\sum_j\:|\langle\psi^{(j)}_k|c^{\dagger}_k |0\rangle|
^{2}\delta\left(\omega-E_k^{(j)}/\hbar\right) \:,
\end{equation}
where $|\psi^{(j)}_k\rangle$ is a complete set of total-quasimomentum $k$
eigenstates of the total Hamiltonian of the coupled e-ph system 
(in our case $H_{\textrm{eff}}$), and $E_k^{(j)}$ the corresponding eigenvalues.
In the simplest case, the above sum over $j$ includes the polaron-ground state $|\psi^{(j=0)}_k\rangle$ 
at quasimomentum $k$ and its attendant continuum of states that correspond to the polaron
with quasimomentum $k-q$ and an unbound phonon with quasimomentum $q$. More generally,
with increasing coupling strength there will be multiple coherent polaron bands 
below the one-phonon continuum (threshold for inelastic scattering)
which sets in at energy $E_{gs}+\hbar\omega_0$~\cite{Engelsberg+Schrieffer:63}, 
where $\omega_0$ is the relevant phonon frequency (in our case $\delta\omega$).
All these coherent (split-off from the one-phonon continuum) polaron states 
contribute to the above sum along with their respective continua.

The spectral function is intimately related to the dynamics of polaron formation.
Let us assume that at $t=0$ a bare-excitation Bloch state with quasimomentum $k$
is prepared and e-ph coupling is turned on (e-ph interaction quench). 
Then $|\psi(0)\rangle=c^{\dagger}_k |0\rangle$ is the state of the 
system at $t=0$, while at a later time $t$ its state is given by 
$|\psi(t)\rangle = \sum_j e^{-\frac{i}{\hbar}E_k^{(j)}t}|\psi_k^{(j)}\rangle
\langle\psi_k^{(j)}|c^{\dagger}_k |0\rangle$. It is straightforward to verify
that by Fourier-transforming the spectral function to the time domain 
we obtain the amplitude $\langle\psi(t)|c^{\dagger}_k |0\rangle$ to 
remain in the initial (bare-excitation) state of the system at time 
$t$~\cite{Ku+Trugman:07}. The corresponding probability $|\langle\psi(t)|
c^{\dagger}_k |0\rangle|^{2}$ yields the quasiparticle residue at time $t$, 
thus describing the dynamics of polaron formation.

In our system, the last procedure can be implemented by making use of the state-preparation protocol 
described in Sec.~\ref{StatePrep} to prepare a bare-excitation Bloch state with quasimomentum $k$
and switching on the qubit-resonator coupling at $t=0$. As a matter of fact, given the character
of the polaron formation in our system -- where at the critical coupling strength
there is an abrupt change from an almost undressed (bare) excitation to a heavilly
dressed small polaron --  a variation of the tunable parameter $\phi_{b0}$ from slightly
below the critical value $(\phi_{b0})_c$ to slightly above this value is essentially
equivalent to an e-ph interaction quench. This should be the most straightforward way 
to implement a quench in our system.

Polaron formation is admittedly a very complex dynamical process even in the 
case of the momentum-independent e-ph coupling~\cite{Emin+Kriman:86,Ku+Trugman:07}. 
Even the very fundamental question related to the time it takes to form a polaron, as well as
a more detailed understanding of how phonon excitations evolve into
the correlated phonon cloud of the polaron quasiparticle, are not fully
answered to date. In the presence of strongly-momentum dependent e-ph interactions, 
which even allow for the occurrence of sharp transitions, it should be even more 
difficult to arrive at a full understanding of this process. Our system, with 
its unique set of experimental tools, should be very useful in this regard.
While the time it takes to form a polaron (after an e-ph interaction quench)
should be possible to extract already from measurements of the average
phonon number (by measuring the photon number on different resonators), 
additional characteristics of the polaron-formation dynamics can be unravelled 
by extracting the spectral function via a Ramsey-interference protocol. 

\section{Summary and Conclusions}\label{Conclusions}
To summarize, we have proposed a superconducting analog quantum simulator
for a model with nonlocal electron-phonon couplings of Su-Schrieffer-Heeger and 
breathing-mode types. The simulator is based on an array of transmon qubits 
and microwave resonators. In this system, the nearest-neighbor qubits are Josephson-coupled 
through an appropriately designed connecting circuit, the latter being also inductively coupled  
to a pair of adjacent resonators. Our setup allows one to simulate the strong excitation-phonon 
coupling regime, characterized by the small-polaron formation, with quite realistic 
values of the circuit parameters. 

The most interesting feature of the investigated model is the occurrence of a 
sharp (first-order) transition at a critical coupling strength. This transition
results from a real level crossing and physically corresponds to the change from a 
nondegenerate single-particle ground state at zero quasimomentum ($K_{\textrm{gs}}=0$) 
to a twofold-degenerate one at nonzero quasimomenta $K_{\textrm{gs}}$ and $-K_{\textrm{gs}}$. 
Aside from the fact that our suggested setup provides a tunable
experimentally platform for observing the sharp transition, what further 
motivated the present work is the circumstance that e-ph coupling in real 
materials is insufficiently strong for observing 
any measurable consequence of this type of transition~\cite{PeierlsOrganics}.

One of the obvious advantages of superconducting Josephson-junction based systems 
compared to other quantum-simulation platforms (trapped ions, polar molecules, Rydberg atoms)
is that they allow realization of strictly nearest-neighbor 
hopping~\cite{Mei+:13}, thus making it possible to simulate relevant polaron
models in a quite realistic fashion. In trapped-ion and polar-molecule 
systems, for instance, the presence of non-nearest neighbor hopping is unavoidable, 
originating from the presence of long-range interactions between their elementary constituents 
(Coulomb interaction between ions and dipolar interaction between polar molecules);
these systems show similar limitations when it comes to mimicking the behavior of 
dispersionless (zero-dimensional) phonons~\cite{Stojanovic+:12,Herrera+Krems:11,Herrera+:13}. 
In addition to these intrinsic advantages of superconducting systems compared to other available
experimental platforms, a unique aspect of our suggested setup is the capability of the
{\em in-situ} altering of the hopping energy through an externally-tunable magnetic flux.

Once experimentally realized, our suggested setup could also be used for studying the 
nonequilibrium aspects of polaron physics, i.e., the dynamics of small-polaron formation.
Experimental studies of such phenomena in traditional solid-state systems are hampered
by the very short dynamical time scales, in addition to a very limited control over 
these systems. Our setup paves the way for a controlled experimental investigation
of this important phenomenon using a Ramsey-interference protocol. It holds 
promise to become an invaluable platform for future studies in this direction.

\acknowledgments
Useful discussions with J. Koch, B. Vlastakis, Z. Minev,
L. I. Glazman, and S. M. Girvin are gratefully acknowledged.
V.M.S. was supported by the SNSF. M.V. was supported by the Serbian Ministry of Science, 
project No. 171027. E.D. acknowledges support from Harvard-MIT CUA, the ARO-MURI 
on Atomtronics, and ARO MURI Quism program. L.T. was supported by NSF-DMR-0956064 
and NSF-CCF-0916303. This research was supported in part 
by NSF PHY11-25915 through a KITP program (V.M.S. and L.T.).

\appendix
\section{Derivation of the Josephson-coupling term in the rotating frame} \label{rotframe_deriv}
In the following, we present derivation of the effective Josephson 
coupling terms in the rotating frame. We will make use of the fact that 
$e^{iH_0t/\hbar}a_{n}e^{-iH_0t/\hbar}=a_{n}e^{-i\omega_0t}$, and that, consequently,
\begin{multline}\label{rotfr_phinres}
e^{\frac{i}{\hbar}H_0t}\:\phi_{n,\textrm{res}}\:e^{-\frac{i}{\hbar}H_0t} \\
=\delta\theta(a_{n+1}e^{-i\omega_0t}+a_{n+1}^{\dagger}
e^{i\omega_0t}-a_{n}e^{-i\omega_0t}-a_{n}^{\dagger}e^{i\omega_0t}) \:,
\end{multline}
and rely on the smallness of $\delta\theta$ and the RWA.
For an arbitrary unitary transformation with a generator $S$ ($S^{\dagger}=-S$) 
applied to an analytic operator function $f(A)$ it holds that 
$e^S\:f(A)\:e^{-S}=f(e^S\:A\:e^{-S})$. Therefore,
\begin{multline}\label{eq:HJ}
e^{\frac{i}{\hbar}H_0t}\:\cos(\phi_{n,\textrm{res}})\:e^{-\frac{i}
{\hbar}H_0t}=\cos\big(\delta\theta(a_{n+1}e^{-i\omega_0t}  \\ 
+a_{n+1}^{\dagger}e^{i\omega_0t}-a_{n}e^{-i\omega_0t}-a_{n}^{\dagger}
e^{i\omega_0t})\big) \:,
\end{multline}
with an analogous relation for $\sin(\phi_{n,\textrm{res}})$.

We start from an expression for the Josephson coupling obtained 
by inserting Eq.~\eqref{eq:ph} into Eq.~\eqref{eq:HJ0}.
By immediately dropping the terms that will be rapidly rotating upon
transformation to the rotating frame, the remaining Josephson 
coupling reads
\begin{multline}\label{eq:HJ}
H_{n}^{J}\approx -\left(2E_{J}\cos\frac{\phi_{n}^{t}}{2}+E_{Jb}\cos\phi_{b0}\right) \\ 
\times\cos\left(\varphi_{n}-\varphi_{n+1}\right)+\mathcal{O}(\delta\theta^2) \:.
\end{multline}
With the aid of Eq.~\eqref{eq:phin}, the term $\cos(\phi_{n}^{t}/2)$
can be written as
\begin{multline}
\cos\frac{\phi_{n}^{t}}{2}=-\sin\left(\frac{\pi
\cos(\omega_{0}t)}{2}\right)\frac{\phi_{n,\textrm{res}}}{2}\\
+\cos\left(\frac{\pi\cos(\omega_{0}t)}{2}\right)+
\mathcal{O}(\delta\theta^2)\:, \label{eq:cosphin}
\end{multline}
where the last two equations have been derived using the 
asymptotic relations $\cos\left(\phi_{n,\textrm{res}}/2\right)=
1+\mathcal{O}(\delta\theta^2)$ and $\sin\left(\phi_{n,\textrm{res}}/2\right)
=\phi_{n,\textrm{res}}/2+\mathcal{O}(\delta\theta^2)$.
At the same time, we will utilize the following well-known expansions
in terms of Bessel functions of the first kind:~\cite{ArfkenWeberBook} 
\begin{multline}
\cos\left(\frac{\pi\cos(\omega_{0}t)}{2}\right)= \\
J_{0}\left(\frac{\pi}{2}\right)-2\sum_{n=1}^{\infty}J_{2n}
\left(\frac{\pi}{2}\right)\cos\left(2n\omega_{0}
t\right)\:,\label{eq:besselcos}
\end{multline}
\begin{multline}
\sin\left(\frac{\pi\cos(\omega_{0}t)}{2}\right)= \\  
2\sum_{n=1}^{\infty}J_{2n-1}\left(\frac{\pi}{2}\right)
(-1)^{n+1}\cos\left((2n-1)\omega_{0}t\right)\:.\label{eq:besselsin}
\end{multline}

We now have to analyze which terms in the last two expansions will remain 
after the transformation to the rotating frame.
The first (time-independent) term in Eq.~\eqref{eq:besselcos} will
clearly be unaffected by this transformation, while the remaining terms
rotate at frequency $2\omega_{0}$ or higher and can therefore be neglected 
by virtue of the RWA. It is also easy to see that the $n=1$ term in Eq.~\eqref{eq:besselsin} 
will also give rise to time-independent terms. Namely, using Eq.~\eqref{rotfr_phinres} 
we straightforwardly obtain
\begin{equation}
\cos\left(\omega_{0}t\right)e^{\frac{i}{\hbar}H_0t}\:\phi_{n,\textrm{res}}
\:e^{-\frac{i}{\hbar}H_0t}=\frac{\phi_{n,\textrm{res}}}{2}+\ldots \:,
\end{equation}
where the ellipses stand for the terms that rotate at frequency $2\omega_{0}$,
and and can therefore be neglected. In this manner, 
after choosing $E_{Jb}=2E_{J}J_{0}(\pi/2)$, we obtain Eq.~\eqref{eq:HJn_final}.
It should be stressed that, being approximately given by Eq.~\eqref{eq:cos}, the
term $\cos\left(\varphi_{n}-\varphi_{n+1}\right)$ is unaffected by the transformation
to the rotating frame.

\end{document}